\documentstyle[aps]{revtex}
\begin{document}
\author{Kaige Wang$^{1,2}$ and De-zhong Cao$^2$}
\address{1.CCAST (World Laboratory), P. O. Box 8730, Beijing 100080\\
2.Department of Physics, Applied Optics Beijing Area Major Laboratory,\\
Beijing Normal University, Beijing 100875, China}
\title{Coincidence Subwavelength Interference by a Classical Thermal Light }
\maketitle

\begin{abstract}
We show that a thermal light random in transverse direction can perform
subwavelength double slit interference in a joint-intensity measurement.
This is the classical version of quantum lithography, and it can be
explained with the correlation of rays instead of the entanglement of
photons.
\end{abstract}

\begin{abstract}
PACS number(s): 42.50.Dv, 42.50.St, 42.25.Hz
\end{abstract}

Noise is usually considered as harmful to communication and interference.
Disorder and chaotic fluctuation may fade information and interference
patterns. Can a noise be used as a signal carrier in communication and a
source in interference? This study is aroused from a recent debate in
quantum optics. In the last decade, the theoretical and experimental studies
have shown that an entangled photon pair generated by the spontaneous
parametric down-conversion (SPDC) exhibits some peculiar effects, such as
sub-wavelength lithography,\cite{yama}-\cite{shi} coincidence or ''ghost''
imaging and interference.\cite{shih1}-\cite{gigi} Indeed, these effects had
not been realized in the classical optics until that time. Especially, the
sub-wavelength interference effect was ever one of the paradoxes relating to
violations of the quantum mechanical uncertainty principle. Recently,
Bennink et al have shown in their experiment that the coincidence imaging
and the coincidence interference can be carried out by a classical light
source.\cite{ben1} Therefore, this demonstrates that the classical
correlation plays the same (or similar) role as the quantum entanglement.
The recent theoretical analysis\cite{gatti} has shown that a thermal or
quasi-thermal source can be charged with such a classical correlation. On
the other hand, the sub-wavelength lithography is considered as a
nonclassical interference surpassing the Rayleigh diffraction limit.
Physically, this effect is explained with the quantum entanglement of photons%
\cite{boto}\cite{shih2} or the photonic de Broglie wavelength of a
multiphoton wavepacket.\cite{yama}\cite{fon1}\cite{fon2} A similar question
has emerged: can the sub-wavelength interference be carried out by a
classical beam, too? In this work, we demonstrate that a thermal light
random in the transverse plane possesses a similar second-order spatial
correlation to that of a photon pair perfectly entangled in the transverse
wavevector. The difference between them is that the former has the
correlation of the wavevector-degeneracy and the latter has the correlation
of the wavevector-conservation. Therefore, a thermal light can perform not
only the ghost imaging and interference as shown in the experiments, but
also the sub-wavelength interference. Due to the different way in
correlation, the observation of the sub-wavelength interference pattern is
taken by a coincidence detection at a pair of symmetric positions. The
result can be use to explain two kinds of macroscopic observation of
sub-wavelength interference occurring in the SPDC of type I crystal.\cite
{kaige}

We consider the first-order and second-order correlations of the fields for
a quantum system consisting of a photon pair and a classical thermal light.
The beam propagates with a central wavevector ${\bf k}_0$ and frequency $%
\omega _0$. For a general two-photon state, the quantum wavefunction in the
Schr\"{o}dinger picture is written as $|\Psi \rangle =\int d{\bf q}_1d{\bf q}%
_2d\omega _1d\omega _2C({\bf q}_1,\omega _1;{\bf q}_2,\omega _2)a_1^{\dagger
}({\bf q}_1,\omega _1)a_2^{\dagger }({\bf q}_2,\omega _2)|0\rangle $, where $%
{\bf q}_i$ and $\omega _i$ ($i=1,2$) are the transverse wavevector (or the
spatial frequency) and the frequency deviation from the central frequency,
respectively. We assume that two photons are distinguishable by their
polarizations. By taking into account the commutation relation of the field
operators, we obtain the first- and second-order spectral correlations for
the two-photon state $|\Psi \rangle $ 
\begin{mathletters}
\label{1}
\begin{eqnarray}
\langle a_1^{\dagger }({\bf q},\omega )a_1({\bf q}^{\prime },\omega ^{\prime
})\rangle &=&\int d{\bf q}_2d\omega _2C^{*}({\bf q},\omega ;{\bf q}_2,\omega
_2)C({\bf q}^{\prime },\omega ^{\prime };{\bf q}_2,\omega _2),  \label{1a} \\
\langle a_2^{\dagger }({\bf q},\omega )a_2({\bf q}^{\prime },\omega ^{\prime
})\rangle &=&\int d{\bf q}_1d\omega _1C^{*}({\bf q}_1,\omega _1;{\bf q}%
,\omega )C({\bf q}_1,\omega _1;{\bf q}^{\prime },\omega ^{\prime }),
\label{1b}
\end{eqnarray}
and 
\end{mathletters}
\begin{equation}
\langle a_1^{\dagger }({\bf q}_1,\omega _1)a_2^{\dagger }({\bf q}_2,\omega
_2)a_2({\bf q}_2^{\prime },\omega _2^{\prime })a_1({\bf q}_1^{\prime
},\omega _1^{\prime })\rangle =C^{*}({\bf q}_1,\omega _1;{\bf q}_2,\omega
_2)C({\bf q}_1^{\prime },\omega _1^{\prime };{\bf q}_2^{\prime },\omega
_2^{\prime }),  \label{2}
\end{equation}
respectively. Equation (\ref{2}) shows explicitly separable product of the
wavefunction and its conjugate one, reflecting the feature of quantum
wavepacket. This implies that the wavevector and frequency correlation may
exist only within the same (positive or negative) frequency component. For
the perfectly entangled two-photon state generated in SPDC, it has $C({\bf q}%
_1,\omega _1;{\bf q}_2,\omega _2)=\delta ({\bf q}_1+{\bf q}_2)\delta (\omega
_1+\omega _2)$ owing to the conservation of momentum and energy in the SPDC
process. Therefore, one obtains 
\begin{equation}
\langle a_i^{\dagger }({\bf q},\omega )a_i({\bf q}^{\prime },\omega ^{\prime
})\rangle =\delta ({\bf q}-{\bf q}^{\prime })\delta (\omega -\omega ^{\prime
}),\qquad (i=1,2)  \label{3}
\end{equation}
\begin{equation}
\langle a_1^{\dagger }({\bf q}_1,\omega _1)a_2^{\dagger }({\bf q}_2,\omega
_2)a_2({\bf q}_2^{\prime },\omega _2^{\prime })a_1({\bf q}_1^{\prime
},\omega _1^{\prime })\rangle =\delta ({\bf q}_1+{\bf q}_2)\delta (\omega
_1+\omega _2)\delta ({\bf q}_1^{\prime }+{\bf q}_2^{\prime })\delta (\omega
_1^{\prime }+\omega _2^{\prime }).  \label{4}
\end{equation}
The results are also true for two photons with the same polarizations.

Then we consider a classical thermal light. We assume a monochromatic plane
wave $E_0\exp [i(k_0z-\omega _0t)]$ illuminating a scattering material
containing disorder scattering centers. After scattering, the field is
written as $E({\bf x},z,t)=\int E({\bf q})\exp [i({\bf q\cdot x}+k_zz-\omega
_0t)]d{\bf q}$ where ${\bf q}$ is the transverse wavevector introduced by
the random scattering and satisfies $|{\bf q}|^2+k_z^2=k_0^2$. Hence, $E(%
{\bf q})$ is the stochastic variable obeying the Gaussian statistics.
However, the scattering waves with different transverse wavevectors are
independent statistically. If $|{\bf q}|<<k_0$, the scattering field is
approximately written as $E({\bf x},z,t)=A({\bf x})\exp [i(k_0z-\omega _0t)]$
where $A({\bf x})=\int E({\bf q})\exp [i{\bf q\cdot x]}d{\bf q}$ is the
slowly varying envelope. As a result, we have defined a monochromatic
thermal light random in both strength and propagation direction. According
to Wiener-Khintchine theorem, the first-order spectral correlation must
satisfy 
\begin{equation}
\langle E^{*}({\bf q})E({\bf q}^{\prime })\rangle =S({\bf q})\delta ({\bf q}-%
{\bf q}^{\prime }),  \label{5}
\end{equation}
where $S({\bf q})$ is the power spectrum of the spatial frequency.
Obviously, Eqs. (\ref{3}) and (\ref{5}) indicate the same incoherence of the
first-order for an ideal two-photon entangled state and a thermal light with
a wide bandwidth. Though the beam is monochromatic, the disorder in spatial
frequency washes out the interference pattern.

For any thermal statistics, the high-order correlation can be represented by
the first-order ones. Hence, the second-order spectral correlation of the
thermal light is written as 
\begin{eqnarray}
&&\langle E^{*}({\bf q}_1)E^{*}({\bf q}_2)E({\bf q}_2^{\prime })E({\bf q}%
_1^{\prime })\rangle   \label{6} \\
&=&\langle E^{*}({\bf q}_1)E({\bf q}_1^{\prime })\rangle \langle E^{*}({\bf q%
}_2)E({\bf q}_2^{\prime })\rangle +\langle E^{*}({\bf q}_1)E({\bf q}%
_2^{\prime })\rangle \langle E^{*}({\bf q}_2)E({\bf q}_1^{\prime })\rangle  
\nonumber \\
&=&S({\bf q}_1)S({\bf q}_2)[\delta ({\bf q}_1-{\bf q}_1^{\prime })\delta (%
{\bf q}_2-{\bf q}_2^{\prime })+\delta ({\bf q}_1-{\bf q}_2^{\prime })\delta (%
{\bf q}_2-{\bf q}_1^{\prime })].  \nonumber
\end{eqnarray}
Different from the case of the two-photon entanglement, in which two photons
with the opposite wavevectors are correlated, Eq. (\ref{6}) shows that two
rays with the degenerate wavevectors are correlated. When the thermal light
is split into two beams in the two arms of a beam splitter, the correlated
degenerate rays are spatially separated. This is the origin of the
coincidence imaging and coincidence interference for a classical source.\cite
{ben1} As a matter of fact, in any coherent beam, there is no such a
correlation between rays, so the coincidence imaging can not occur.

Now, we discuss the double slit interference in a joint-intensity
observation. Let a beam illuminate a double slit with the slit-width $b$ and
the distance between two slits $d$. The double slit and the detection plane
are placed at the two focal planes of a lens with the focal length $f$. The
Fourier transform of the double-slit function is written as $\widetilde{T}%
(q)=(2b/\sqrt{2\pi })\sin $c$(qb/2)\cos (qd/2)$. For simplicity, we omit the
time variable. This can be done by using a narrow frequency filter in beam.
According to Ref.\cite{kaige}, the second-order spatial correlation of the
field in the detection plane is obtained to be 
\begin{eqnarray}
&&\langle E_d^{*}(x_1)E_d^{*}(x_2)E_d(x_2)E_d(x_1)\rangle   \label{7} \\
&=&\frac{k_0^2}{(2\pi f)^2}\int \widetilde{T}(\frac{k_0x_1}f-q_1)\widetilde{T%
}(\frac{k_0x_2}f-q_2)\widetilde{T}(\frac{k_0x_2}f-q_2^{\prime })\widetilde{T}%
(\frac{k_0x_1}f-q_1^{\prime })  \nonumber \\
&&\times \langle E^{*}(q_1)E^{*}(q_2)E(q_2^{\prime })E(q_1^{\prime })\rangle
dq_1dq_2dq_2^{\prime }dq_1^{\prime }.  \nonumber
\end{eqnarray}
For an ideal two-photon entangled state, by using Eq. (\ref{4}), one has $%
\langle E_d^{*}(x_1)E_d^{*}(x_2)E_d(x_2)E_d(x_1)\rangle \propto \widetilde{T}%
^2[\frac{k_0}f(x_1+x_2)].$ If one uses a two-photon detector to scan the
position, the sub-wavelength interference fringe with the perfect visibility
can be observed. Since the two-photon detector is not available at present,
in the experiment, the two-photon detection is carried out by a coincidence
measurement (CM) of two orthogonally polarized photons.\cite{fon1}\cite
{shih2} However, for a thermal light described above, one obtains 
\begin{eqnarray}
&&\langle E_d^{*}(x_1)E_d^{*}(x_2)E_d(x_2)E_d(x_1)\rangle   \label{9} \\
&=&\frac{k_0^2}{(2\pi f)^2}\left\{ \int \widetilde{T}^2(\frac{k_0x_1}f%
-q)S(q)dq\int \widetilde{T}^2(\frac{k_0x_2}f-q)S(q)dq+\left[ \int \widetilde{%
T}(\frac{k_0x_1}f-q)\widetilde{T}(\frac{k_0x_2}f-q)S(q)dq\right] ^2\right\} .
\nonumber
\end{eqnarray}
In the broadband limit, we set $S(q)\approx S(0)$ and obtain approximately 
\begin{equation}
\langle E_d^{*}(x_1)E_d^{*}(x_2)E_d(x_2)E_d(x_1)\rangle =\frac{k_0^2S^2(0)}{%
2\pi f^2}\left\{ \widetilde{T}^2(0)+\widetilde{T}^2[\frac{k_0}f%
(x_1-x_2)]\right\} .  \label{10}
\end{equation}
When two detectors are placed in a pair of symmetric positions, $x_1=-x_2=x$%
, to perform a joint-intensity measurement, one observes the sub-wavelength
interference fringe with the 50\% visibility. In the general case, we set
the Gaussian spectrum $S(q)=(\sqrt{2\pi }w)^{-1}\exp [-q^2/(2w^2)]$. Figure
1 shows the interference fringe as function of the normalized bandwidth $%
W=wb/(2\pi )$. At very small bandwidth, the interference fringe is
''normal'', i.e. the same as that for a coherent beam. As the bandwidth is
increasing, the sub-wavelength interference effect is appearing and growing.

Physically, the difference of the sub-wavelength interference effects for
the two-photon entangled state and the thermal light comes from the manners
of the correlation, that is, the wavevector conservation and the wavevector
degeneracy. In Fig. 2, we illustrate the mechanism of the sub-wavelength
interference effect of the classical thermal correlation. Due to the
correlation of wavevector degeneracy, a pair of correlated rays $E(q)$ and $%
E^{*}(q)$ interfere with another pair of correlated rays $E(q^{\prime })$
and $E^{*}(q^{\prime })$, resulting in twice the optical path difference.

The beam generated in SPDC may incorporate both the quantum entanglement and
the classical thermal correlation. A plane-wave pump field activates a $\chi
^{(2)}$ nonlinear crystal, the basic unitary transformation is described by%
\cite{gigi} 
\begin{equation}
a_m({\bf q},\omega )=U_m({\bf q},\omega )a_m^{in}({\bf q},\omega )+V_m({\bf q%
},\omega )a_n^{in\dagger }(-{\bf q},-\omega )\qquad (m\neq n=s,i),
\label{11}
\end{equation}
where $a_m({\bf q},\omega )$ and $a_m^{in}({\bf q},\omega )$ are the output
and input field operators, respectively. The first-order correlation is
obtained to be 
\begin{equation}
\langle a_m^{\dagger }({\bf q},\omega )a_m({\bf q}^{\prime },\omega ^{\prime
})\rangle =|V_m({\bf q},\omega )|^2\delta ({\bf q}-{\bf q}^{\prime })\delta
(\omega -\omega ^{\prime }).  \label{12}
\end{equation}
For type I crystal, one may omit the subscript in Eq. (\ref{11}) and the
second-order correlation is written as 
\begin{eqnarray}
&&\langle a^{\dagger }({\bf q}_1,\omega _1)a^{\dagger }({\bf q}_2,\omega
_2)a({\bf q}_2^{\prime },\omega _2^{\prime })a({\bf q}_1^{\prime },\omega
_1^{\prime })\rangle  \label{13} \\
&=&V^{*}({\bf q}_1,\omega _1)V({\bf q}_1^{\prime },\omega _1^{\prime
})U^{*}(-{\bf q}_1,-\omega _1)U(-{\bf q}_1^{\prime },-\omega _1^{\prime
})\delta ({\bf q}_1+{\bf q}_2)\delta (\omega _1+\omega _2)\delta ({\bf q}%
_1^{\prime }+{\bf q}_2^{\prime })\delta (\omega _1^{\prime }+\omega
_2^{\prime })  \nonumber \\
&&+|V({\bf q}_1,\omega _1)V({\bf q}_2,\omega _2)|^2[\delta ({\bf q}_1-{\bf q}%
_1^{\prime })\delta ({\bf q}_2-{\bf q}_2^{\prime })\delta (\omega _1-\omega
_1^{\prime })\delta (\omega _2-\omega _2^{\prime })+\delta ({\bf q}_1-{\bf q}%
_2^{\prime })\delta ({\bf q}_2-{\bf q}_1^{\prime })\delta (\omega _1-\omega
_2^{\prime })\delta (\omega _2-\omega _1^{\prime })].  \nonumber
\end{eqnarray}
The first term shows the same correlation as the two-photon entangled state,
whereas the second term shows the same correlation as the thermal light.
Note that this result is valid for general coupling gain of SPDC, even if
the converted beam contains a large number of photons. When the gain is very
small, the second term is negligible in comparison with the first term and
the converted field is approximately in a two-photon entangled state. As the
gain is increasing, two kinds of macroscopic observation of the
sub-wavelength interference co-exist.\cite{kaige} For type II crystal,
however, the second-order correlation of two orthogonally polarized beams is
obtained to be 
\begin{eqnarray}
&&\langle a_m^{\dagger }({\bf q}_1,\omega _1)a_n^{\dagger }({\bf q}_2,\omega
_2)a_n({\bf q}_2^{\prime },\omega _2^{\prime })a_m({\bf q}_1^{\prime
},\omega _1^{\prime })\rangle \qquad (m\neq n=s,i)  \label{14} \\
&=&V_m^{*}({\bf q}_1,\omega _1)V_m({\bf q}_1^{\prime },\omega _1^{\prime
})U_n^{*}(-{\bf q}_1,-\omega _1)U_n(-{\bf q}_1^{\prime },-\omega _1^{\prime
})\delta ({\bf q}_1+{\bf q}_2)\delta (\omega _1+\omega _2)\delta ({\bf q}%
_1^{\prime }+{\bf q}_2^{\prime })\delta (\omega _1^{\prime }+\omega
_2^{\prime })  \nonumber \\
&&+|V_m({\bf q}_1,\omega _1)V_n({\bf q}_2,\omega _2)|^2\delta ({\bf q}_1-%
{\bf q}_1^{\prime })\delta ({\bf q}_2-{\bf q}_2^{\prime })\delta (\omega
_1-\omega _1^{\prime })\delta (\omega _2-\omega _2^{\prime }).  \nonumber
\end{eqnarray}
The first term is the same as that for type I case, but the second term does
not represent a thermal correlation. One can observe the quantum
subwavelength interference only.\cite{kaige} The thermal correlation in
type-II crystal can be recovered as long as the signal or idler beam is
extracted by a polarization beamsplitter. The second-order correlation for
the signal/idler beam reads 
\begin{eqnarray}
&&\langle a_m^{\dagger }({\bf q}_1,\omega _1)a_m^{\dagger }({\bf q}_2,\omega
_2)a_m({\bf q}_2^{\prime },\omega _2^{\prime })a_m({\bf q}_1^{\prime
},\omega _1^{\prime })\rangle \qquad (m=s,i)  \label{15} \\
&=&|V_m({\bf q}_1,\omega _1)V_n({\bf q}_2,\omega _2)|^2[\delta ({\bf q}_1-%
{\bf q}_1^{\prime })\delta ({\bf q}_2-{\bf q}_2^{\prime })\delta (\omega
_1-\omega _1^{\prime })\delta (\omega _2-\omega _2^{\prime })+\delta ({\bf q}%
_1-{\bf q}_2^{\prime })\delta ({\bf q}_2-{\bf q}_1^{\prime })\delta (\omega
_1-\omega _2^{\prime })\delta (\omega _2-\omega _1^{\prime })].  \nonumber
\end{eqnarray}
We obtain the thermal correlation for either signal or idler beam.

In conclusion, we show a classical version of quantum subwavelength
lithography. This demonstrates that the classical correlation of a thermal
light plays the similar role as the quantum entanglement of two photons in
the optical subwavelength interference, in addition to the coincidence
imaging and interference.

This research was supported by the National Fundamental Research Program of
China with No. 2001CB309310, and the National Natural Science Foundation of
China, Project Nos. 60278021 and 10074008.

Captions of Figures

Fig. 1 Coincidence interference patterns for the different normalized
bandwidths $W=wb/(2\pi )$ of spatial frequency spectrum of a thermal light. $%
X=xk_0b/(2\pi f)$ is the normalized position in the detection plane and the
double slit parameter is taken as $d=4b$.

Fig. 2 (a) Schematic diagram of coincidence subwavelength interference for a
thermal light. (b) The right and left sides show respectively the field $%
E(q) $ and its conjugate $E^{*}(q)$ to be correlated, resulting in twice the
optical path difference.

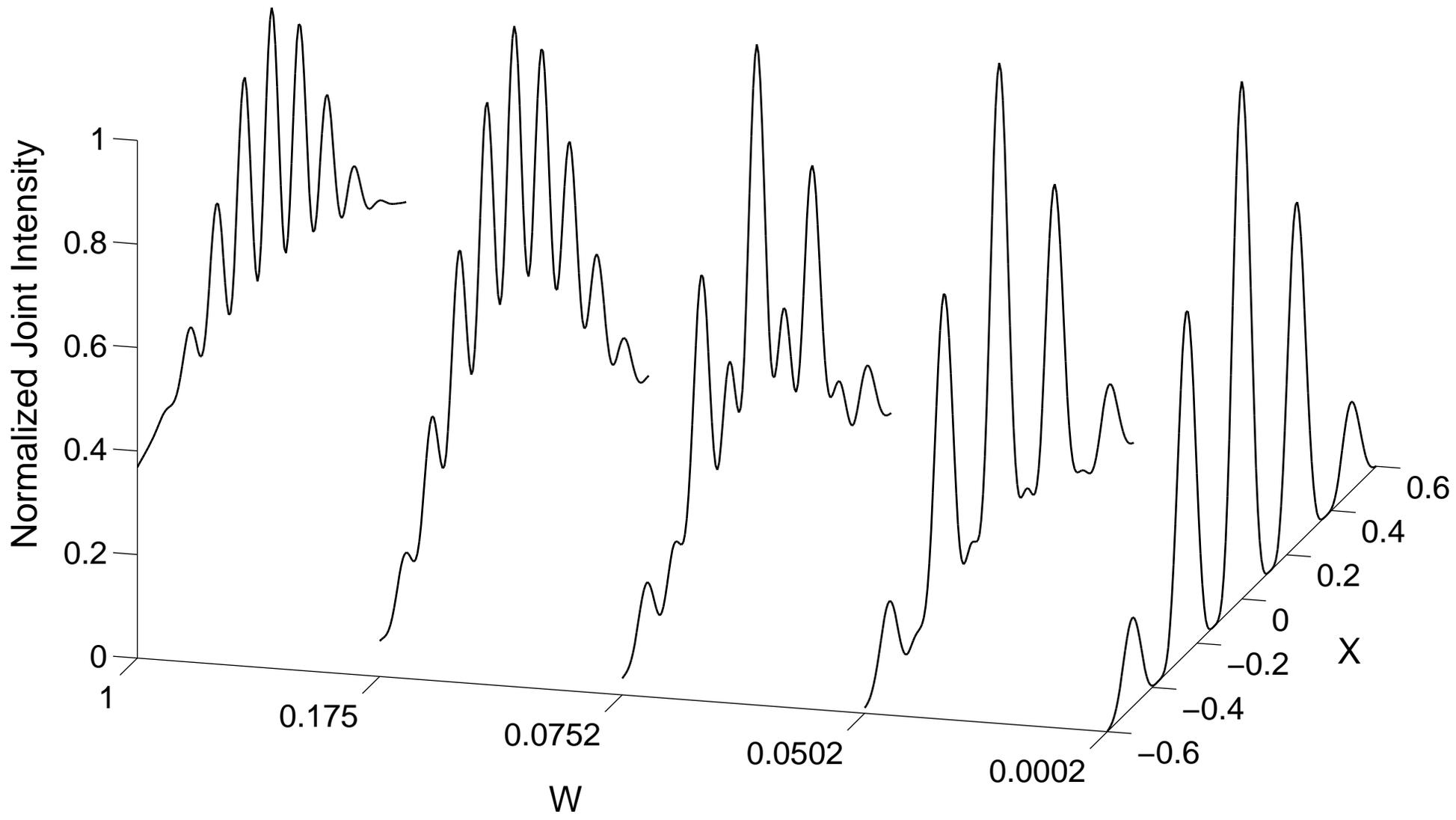
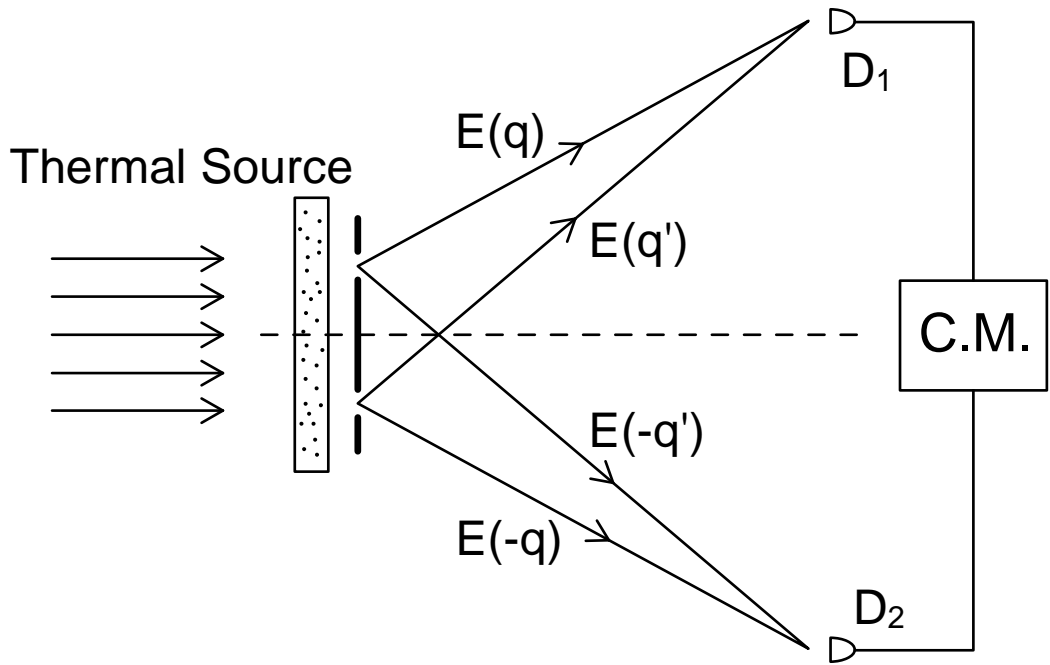
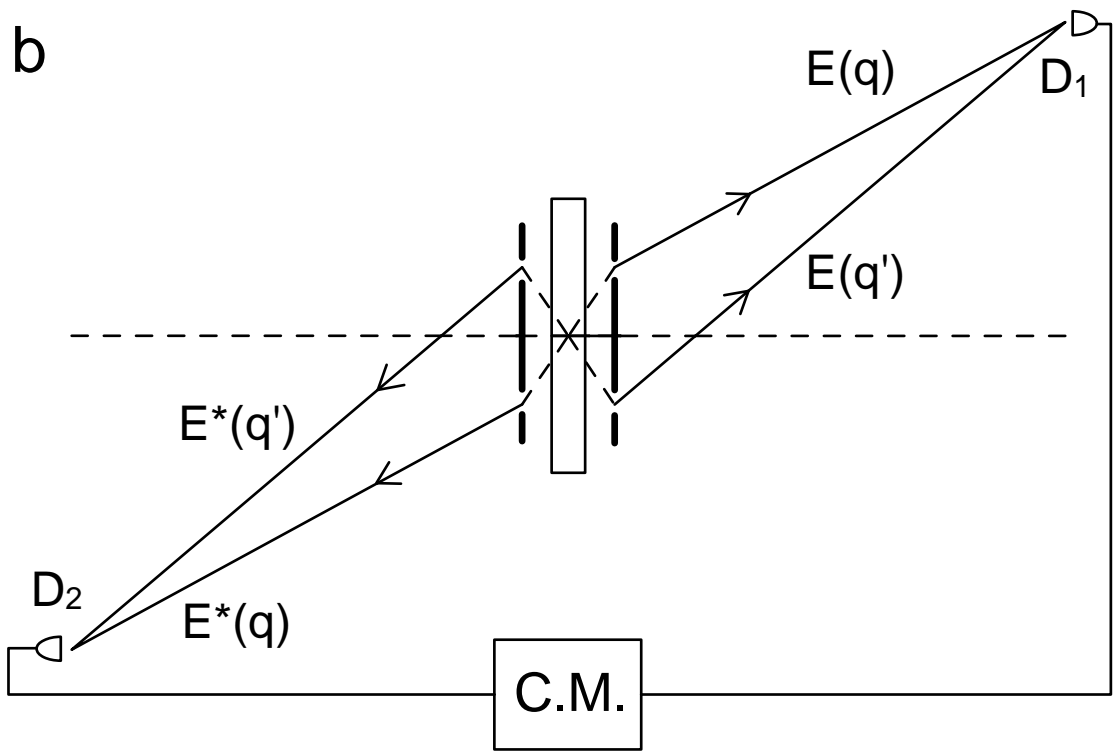
\end{document}